\documentclass[aps,prd,onecolumn,superscriptaddress]{revtex4-2}
\usepackage{graphicx}
\usepackage{amssymb}  
\usepackage{mathtools}

\begin{document}

\title{Exact Solutions for a Teleparallel Cosmological Model with Vector Field via Noether Symmetry}

\author{Yusuf Kucukakca}
\email{ykucukakca@akdeniz.edu.tr}
\affiliation{Department of Physics, Faculty of Science, Akdeniz University, 07058 Antalya, Turkey}

\author{Fereshteh Omidvar}
\email{fereshteh.omvr@gmail.com}
\affiliation{Faculty of Physics, University of Tabriz, Tabriz, 51666-16471, Iran}

\author{Amin Rezaei Akbarieh}
\email{am.rezaei@tabrizu.ac.ir}
\affiliation{Faculty of Physics, University of Tabriz, Tabriz, 51666-16471, Iran}
\affiliation{Department of Mathematics, Faculty of Sciences, Bilkent University, 06800 Ankara, Turkey}

\date{August 2025}

\begin{abstract}
We study a cosmological model in the framework of teleparallel gravity, where a vector field $A_\mu$ is non-minimally coupled to the torsion scalar $T$ in a flat Friedmann-Robertson-Walker (FRW) universe. Using the Noether symmetry approach, we identify specific forms for the coupling function $g(\xi)$ and the potential  $V(\xi) = V_0 \xi^n$, with $\xi \equiv A_\mu A^\mu$. The method allows us to find exact analytical solutions for the scale factor $a(t)$ and the scalar function $\xi(t)$. General solutions for arbitrary values of $n$ are derived, but special cases such as $n = 1$ and $n = 3$ are studied separately due to their distinct behavior. For $n = 1$, the model describes a transition from decelerated to accelerated expansion, and depending on the value of the model parameter, a transition to a phantom phase is also possible, which is consistent with phantom dark energy. For the special case $n=3$, this model first experiences a deceleration phase and then enters a stable accelerating phase, such that the acceleration parameter $q(t)$ changes from positive to negative values and the equation of state parameter $\omega_\xi(t)$ tends to $-1$ at late times, similar to quintessence dark energy. Unlike the case $n=1$, in this case, no transition to the phantom phase is observed. Therefore, this model can describe the behavior of the universe during periods of dark energy dominance.
\end{abstract}

\maketitle
\section{Introduction}
\label{int}

The discovery of the universe's accelerated expansion, first evidenced by observations of type Ia supernovae and later corroborated by cosmic microwave background (CMB) radiation and large-scale structure surveys \cite{SupernovaSearchTeam:1998fmf,SupernovaCosmologyProject:1998vns,WMAP:2003elm}, has fundamentally transformed our understanding of cosmology. These observations indicate that the universe is not only expanding but doing so at an accelerating rate, a phenomenon attributed to a mysterious component known as dark energy, which constitutes approximately 68\% of the universe's energy budget \cite{Planck:2015fie,Planck:2018vyg}. General Relativity (GR), the cornerstone of modern gravitational theory, successfully describes a wide range of gravitational phenomena, from planetary orbits to gravitational waves. However, explaining cosmic acceleration within GR requires the introduction of a cosmological constant or an exotic form of energy with negative pressure, the physical origin of which remains elusive \cite{Peebles:2002gy}. Moreover, GR faces challenges in addressing other cosmological puzzles, such as the nature of dark matter, the mechanisms of early-time inflation, and the resolution of singularities in extreme gravitational regimes \cite{Clifton:2011jh,Ishak:2018his}. These shortcomings have motivated extensive efforts to develop alternative gravitational theories and cosmological models to account for the observed dynamics of the universe.

Various theoretical frameworks, broadly categorized into two approaches: modifications to the matter-energy content of the universe and modifications to the gravitational framework itself \cite{Peebles:2002gy,Yoo:2012ug,Patil:2023rqy}. The simplest model for dark energy is the cosmological constant, introduced by Einstein, which provides a constant energy density driving the accelerated expansion \cite{Bousso:2007gp}. However, the cosmological constant suffers from the fine-tuning problem \cite{Diaz-Pachon:2024nsq}, as its observed value is orders of magnitude smaller than theoretical predictions from quantum field theory \cite{Gubitosi:2012hu}. To address this, dynamical dark energy models have been proposed \cite{Avsajanishvili:2023jcl,Copeland:2006wr}, primarily involving scalar fields \cite{Ratra:1987rm,Wetterich:1987fm}. The quintessence model, characterized by a canonical scalar field with a slowly varying potential, allows the equation of state parameter to evolve, potentially mimicking dark energy behavior \cite{Caldwell:1997ii,Steinhardt:1999nw}. Variants of scalar field models include k-essence, which introduces non-canonical kinetic terms to drive acceleration \cite{Rozas-Fernandez:2011fdk}, phantom fields with negative kinetic energy leading to an equation of state parameter less than \(-1\) \cite{Ludwick:2017tox}, and tachyon fields inspired by string theory \cite{Calcagni:2006ge}. Additionally, scalar-tensor theories, such as Brans-Dicke gravity, couple a scalar field to the curvature scalar, modifying the gravitational dynamics \cite{Basumatary:2024cak,Sobhanbabu:2022woo}. While these models offer flexibility in describing cosmic acceleration, they often require fine-tuning of parameters and face challenges in aligning with observational constraints across all cosmological epochs \cite{WMAP:2012nax}.

Beyond scalar field models, alternative approaches to dark energy involve modifying the gravitational sector. Modified gravity theories, such as \( f(R) \) gravity, extend the Einstein-Hilbert action by introducing a general function of the Ricci scalar, enabling late-time acceleration without exotic matter \cite{Nojiri:2003ft}. Other frameworks include Gauss-Bonnet gravity, which incorporates higher-order curvature terms \cite{Nojiri:2005vv}, and massive gravity, which assigns a mass to the graviton \cite{deRham:2010kj}. Despite their theoretical appeal, these models often introduce mathematical complexities and require rigorous testing against observational data, such as CMB anisotropies and supernova distance moduli \cite{SupernovaCosmologyProject:2011ycw}. The diversity of these approaches underscores the ongoing challenge of identifying a consistent model for dark energy that is both theoretically robust and observationally viable.

Among modified gravity theories, teleparallel gravity has emerged as a promising framework by reformulating gravitation in terms of torsion rather than curvature \cite{Bahamonde:2021gfp,Cai:2015emx}. In teleparallel gravity, the dynamical variables are represented by tetrad fields, and spacetime is described by a torsion scalar, which has a role similar to the Ricci scalar in General Relativity theory \cite{Aldrovandi:2013wha}. The teleparallel equivalent of general relativity (TEGR) can be generalized to \( f(T) \) gravity, where the Lagrangian is an arbitrary function of \( T \), offering a versatile platform to model both early-time inflation and late-time cosmic acceleration \cite{Capozziello:2015rda,Myrzakulov:2012sp}.  It should also be emphasized that the equations of this theory result in second-order field equations that are simpler than equations of higher-order theories such as $f(R)$ gravity. Recent extensions of teleparallel gravity incorporate additional fields, such as scalar or vector fields, to explore the interplay between geometry and matter, potentially addressing the shortcomings of GR in explaining dark energy \cite{Kucukakca:2022bos,Amiri:2024unw,Kucukakca:2024eoq}. Unlike scalar fields, vector fields introduce directional degrees of freedom, which can lead to novel cosmological behaviors, such as anisotropic effects or dark energy-like dynamics, making them particularly intriguing for cosmological applications \cite{Coelho:2025vmo,Armendariz-Picon:2004say,Nakamura:2019phn}.

Although the symmetry approach plays a crucial role in analyzing physical problems, Noether symmetry is particularly significant due to its advantage in identifying the conservation laws of dynamical systems. The existence of conserved quantities derived from Noether symmetry enables the integrability of dynamical systems obtained from the variational principle and the Noether symmetry vector can be used to define a coordinate transformation that introduces a cyclic variable in the generalized coordinates In addition, the existence of this symmetry provides insight into the possible forms of the unknown functions in the dynamical system. In this context, this approach can be considered as a criterion for selecting a physical model. Noether symmetry has been widely applied in both theoretical and mathematical physics for nearly a century, and in recent years, it has become a crucial tool for obtaining exact solutions and constructing physically valid models in cosmology and modified gravitational theories. In this context, it has been studied, for example, in scalar field theories \cite{Capozziello:2009te,Camci:2007zz,Kucukakca:2012zm,Paliathanasis:2023dfz,Kucukakca:2020ydp,Laya:2023ibq,Oz:2017jor,Kucukakca:2020kuh}, in gravity theories involving fermionic \cite{Kucukakca:2014vja,Gecim:2014iza} and tachyonic fields \cite{Motavalli:2016gid,Akbarieh:2023nvj}, in the $f(R)$ gravity theory \cite{Capozziello:2008ch,Kucuakca:2011np,Kucukakca:2016mvs} and in modified teleparallel gravity \cite{Wei:2011aa,Atazadeh:2011aa,Tajahmad:2017ywa,Kucukakca:2013mya,Gecim:2017hmn,Bahamonde:2016grb,Kucukakca:2018twx,Paliathanasis:2022mcc,Duchaniya:2022hiy}.

In this study, we investigate a teleparallel cosmological model where the gravitational action includes a vector field \( A_\mu \) non-minimally coupled to the torsion scalar \( T \). The action is defined with a coupling function \( f(\xi) \), where \( \xi = A_\mu A^\mu \), and a potential \( V(\xi) = V_0 \xi^n \), generalizing TEGR to incorporate the dynamics of the vector field. To ensure compatibility with the homogeneity and isotropy of the universe, we adopt a flat Friedmann-Lema\^{\i}tre-Robertson-Walker (FLRW) spacetime and a spatially isotropic vector field configuration. The inclusion of the vector field introduces additional complexity to the field equations, as its dynamics are governed by both the teleparallel geometry and the field strength tensor \( F_{\mu\nu} \). To address this, we employ the Noether symmetry approach, a powerful method that leverages symmetries of the Lagrangian to identify conserved quantities and simplify the dynamical system. By applying Noether's theorem to the point-like Lagrangian, we constrain the coupling function \( g(\xi) \) (derived from \( f(\xi) \)) and the potential \( V(\xi) \), and change of a coordinate transformation that renders the system integrable, enabling the derivation of exact solutions for the scale factor \( a(t) \) and the scalar field \( \xi(t) \).

Our analysis focuses on general solutions for arbitrary values of the parameter $n$, with particular emphasis on the important physical state $n=1$, which describes a universe that transitions from a decelerating phase to a phase with positive acceleration; in this case, the equation of state parameter $\omega_\xi(t)$ gradually approaches the value $-1$, and depending on the value of the model parameter, it can even enter a phantom phase, which is consistent with some observations of phantom dark energy. In contrast, for the special case $n=3$, the model also exhibits a transition from a decelerating to an accelerating phase, but unlike the case $n=1$, no transition to a phantom phase occurs there, and the cosmological parameters eventually converge to a stable behavior similar to quintessence dark energy. The parameters obtained for Hubble and acceleration provide a clear view of the expansion dynamics and its observable signatures. This study, by examining the role of vector fields in the framework of teleparallel gravity and demonstrating the effectiveness of the Noether symmetry approach in simplifying and solving complex cosmological systems, is an effective step towards a better understanding of dark energy.

The article is organized as follows: Section 2 outlines the Noether symmetry approach, including the derivation of cyclic variable and exact solutions. Section 3 examines special cases for \( n \), analyzing their cosmological implications through parameters such as the Hubble and deceleration parameters. Section 4 concludes with a summary of the findings and discusses future directions, including potential extensions to the model and numerical validations against observational datasets.

\section{Introduction to the Model}
\label{sec:model_intro}

In this study, we explore a cosmological model within the framework of teleparallel gravity, where the gravitational interaction is described by the torsion scalar rather than the curvature scalar of general relativity. The action governing our model is given by
\begin{equation}
\label{eq:action}
S = \int d^4 x \sqrt{-g} \left[ f(\xi) T - \frac{1}{4} F_{\mu\nu} F^{\mu\nu} - V(\xi) \right],
\end{equation}
where \( T \) is the torsion scalar, \( F_{\mu\nu} = \partial_\mu A_\nu - \partial_\nu A_\mu \) is the field strength tensor of the vector field \( A_\mu \), and \( V(\xi) \) is the potential of the scalar field \( \xi \equiv A_\mu A^\mu \). The function \( f(\xi) \) represents an arbitrary coupling function between \( \xi \) and the torsion scalar \( T \), generalizing the teleparallel equivalent of general relativity (TEGR). Teleparallel gravity provides an alternative formulation of gravitation, where the dynamics are described by the tetrad fields \( e^a_\mu \) and the Weitzenböck connection has vanishing curvature but non-zero torsion \cite{Aldrovandi:2013wha}. The metric is related to the Minkowski metric via \( g_{\mu\nu} = \eta_{ab} e^a_\mu e^b_\nu \), and the tetrads satisfy orthonormality conditions \( e^a_\mu e^\mu_b = \delta^a_b \) and \( e^a_\mu e^\nu_a = \delta^\nu_\mu \). The torsion tensor is defined as \( T^\rho_{\mu\nu} = \Gamma^\rho_{\nu\mu} - \Gamma^\rho_{\mu\nu} \), where \( \Gamma^\rho_{\mu\nu} = e^\rho_a \partial_\mu e^a_\nu \) is the Weitzenböck connection. The super-potential is given by
\begin{equation}
S_\rho^{\mu\nu} = \frac{1}{2} \left( K^{\mu\nu}_\rho + \delta^\mu_\rho T^\sigma_\sigma{}^\nu - \delta^\nu_\rho T^\sigma_\sigma{}^\mu \right),
\end{equation}
where \( K^{\mu\nu}_\rho = \frac{1}{2} \left( T^\nu{}_\rho{}^\mu + T^\mu{}_\rho{}^\nu - T_\rho{}^{\mu\nu} \right) \) is the contorsion tensor. The torsion scalar \( T \), defined as
\begin{equation}
T = S_\rho{}^{\mu\nu} T^\rho_{\mu\nu},
\end{equation}
encapsulates the gravitational dynamics. In the covariant formulation of teleparallel gravity, the spin connection \( \omega^a_{b\mu} \) accounts for inertial effects and ensures Lorentz invariance of the field equations \cite{Kucukakca:2022bos}. In the Weitzenböck gauge, the spin connection can be set to zero in specific frames, simplifying calculations without loss of generality. This framework has been shown to describe both early-time inflation and late-time accelerated expansion, making it a compelling candidate for modeling dark energy \cite{Kucukakca:2022bos}.

For cosmological applications, we adopt the flat Friedmann-Robertson-Walker (FRW) metric,
\begin{equation}
ds^2 = -dt^2 + a(t)^2 \left( dx^2 + dy^2 + dz^2 \right),
\end{equation}
with the tetrad $e^{a}_{\mu}=diag(1,a,a,a)$, which is compatible with the Weitzenböck gauge. The torsion scalar for this metric is \( T = -6 H^2 \), where \( H = \dot{a}/a \) is the Hubble parameter, and dots denote derivatives with respect to cosmic time \( t \).
To ensure compatibility with the homogeneity and isotropy of the universe, the vector field is chosen as
\begin{equation}
\label{eq:vector_field}
A_\mu = (0, a(t) \phi(t), a(t) \phi(t), a(t) \phi(t)),
\end{equation}
where \( \phi(t) \) is a time-dependent scalar field, and the equal spatial components maintain isotropy. Using \ref{eq:vector_field}, we derive the physical quantities
\begin{equation}
\label{eq:xi}
\xi = A_\mu A^\mu = 3 \phi^2,
\end{equation}
\begin{equation}
\label{eq:maxwell_tensor}
F_{\mu\nu} F^{\mu\nu} = -6 \left( \dot{\phi}^2 + 2 H \phi \dot{\phi} + H^2 \phi^2 \right).
\end{equation}
The field equation for the vector field is satisfied by this choice, consistent with the cosmological principle.

The point-like Lagrangian is derived in terms of the scale factor \( a(t) \) and the scalar field \( \xi(t) \) after integrating over spatial coordinates
\begin{equation}
\label{eq:lagrangian}
\mathcal{L} = \frac{1}{2} g(\xi) \dot{a}^2 a + \frac{1}{8} a^3 \frac{\dot{\xi}^2}{\xi} + \frac{1}{2} a^2 \dot{a} \dot{\xi} - a^3 V(\xi),
\end{equation}
where \(g(\xi)\equiv\xi-12f(\xi)\), and primes (e.g., \( g'(\xi) \), \( V'(\xi) \)) denote derivatives with respect to \( \xi \). The equations of motion are obtained by varying the Lagrangian with respect to \( a(t) \) and \( \xi(t) \). The Euler-Lagrange equation for \( a(t) \) yields:
\begin{equation}
\label{eq:eom_a}
 4 g(\xi) \dot{a}^2 + 8 a \left( \dot{a} g'(\xi) \dot{\xi} + g(\xi) \ddot{a} \right) - a^2 \left( -24 V(\xi) + \frac{3 \dot{\xi}^2}{\xi} - 4 \ddot{\xi} \right)  = 0.
\end{equation}
For \( \xi(t) \), after simplifying, the equation of motion is
\begin{equation}
\label{eq:eom_xi}
a^2 \dot{\xi}^2 + 4 \xi^2 \left[ \dot{a}^2 (-2 + g'(\xi)) - a \left( 2 a V'(\xi) + \ddot{a} \right) \right] - 2 a \xi \left( 3 \dot{a} \dot{\xi} + a \ddot{\xi} \right) = 0.
\end{equation}
The energy function, defined as \( E = \dot{a} \frac{\partial \mathcal{L}}{\partial \dot{a}} + \dot{\xi} \frac{\partial \mathcal{L}}{\partial \dot{\xi}} - \mathcal{L} \), is:
\begin{equation}
\label{eq:energy}
E = \frac{1}{8} a \left[ 4 g(\xi) \dot{a}^2 + 4 a \dot{a} \dot{\xi} + a^2 \left( 8 V(\xi) + \frac{\dot{\xi}^2}{\xi} \right) \right].
\end{equation}
Setting \( E = 0 \), as justified by the homogeneity and isotropy of the FRW spacetime, we obtain the Friedmann-like equation
\begin{equation}
\label{eq:friedmann}
H^2 = \frac{1}{4 g(\xi)} \left[ 8 V(\xi) + \frac{\dot{\xi}^2}{\xi} + 4 H \dot{\xi} \right] \equiv \frac{\rho_\xi}{6g(\xi)},
\end{equation}
where the energy density of the scalar field is defined as
\begin{equation}
\label{eq:rho}
\rho_\xi = \frac{3}{2} \left[ 8 V(\xi) + \frac{\dot{\xi}^2}{\xi} + 4 H \dot{\xi} \right].
\end{equation}
To derive the pressure \( p_\xi \), we use the second Friedmann equation, obtain by rewriting Eq. (\ref{eq:eom_a}) in terms of \( H = \dot{a}/a \), \( \ddot{a} = (\dot{H} + H^2) a \)
\begin{equation}
\label{eq:friedmann2}
2\dot{H} + 3 H^2 = \frac{1}{8 f(\xi)} \left[ 24 V(\xi) - \frac{3 \dot{\xi}^2}{\xi} + 4 \ddot{\xi} - 8 f'(\xi) H \dot{\xi} \right].
\end{equation}
The pressure \( p_\xi \) is then derived from the standard form of the second Friedmann equation, \( 2\dot{H} + 3 H^2 = -\frac{p_\xi}{2g(\xi)} \), yielding
\begin{equation}
\label{eq:pressure}
p_\xi = \frac{1}{4} \left[ -24 V(\xi) + \frac{3 \dot{\xi}^2}{\xi} - 4 \ddot{\xi} + 8 f'(\xi) H \dot{\xi} \right].
\end{equation}
The equation of state parameter for the scalar field is defined as
\begin{equation}
\label{eq:omega}
\omega_\xi = \frac{p_\xi}{\rho_\xi}.
\end{equation}
The evolution of equation of state parameter \( \omega \) is critical for understanding the cosmological phases, such as matter-dominated (\( \omega =0 \)), radiation-dominated (\( \omega = 1/3 \)), or dark energy-dominated (\( \omega = -1 \)) eras, depending on the forms of \( f(\xi) \) and \( V(\xi) \).

\section{Noether Symmetry Approach}
\label{sec:noether}

The Noether symmetry approach is a powerful method in cosmology for identifying symmetries in the Lagrangian that lead to conserved quantities, thereby simplifying the dynamics of complex systems \cite{Capozziello:1996bi}. According to Noether's theorem, every differentiable symmetry of the action corresponds to a conserved quantity, which can be used to reduce the order of the dynamical system and, in some cases, yield exact solutions \cite{Capozziello:1996bi}. In the context of our vector field model, we apply this approach to the point-like Lagrangian \ref{eq:lagrangian} to determine the forms of the coupling function \( f(\xi) \) and the potential \( V(\xi) \), and to find analytical solutions for the scale factor \( a(t) \) and the scalar field \( \xi(t) \). The presence of Noether symmetries not only constrains the functional forms of \( f(\xi) \) and \( V(\xi) \), but also facilitates the identification of a cyclic variable that render the system integrable \cite{Capozziello:1996bi}.

\subsection{Noether Symmetry Conditions}
\label{subsec:noether_conditions}

To apply the Noether symmetry approach, we define the symmetry generator as
\begin{equation}
\label{eq:symmetry_generator}
X = \alpha(a, \xi) \frac{\partial}{\partial a} + \beta(a, \xi) \frac{\partial}{\partial \xi} + \dot{\alpha}(a, \xi) \frac{\partial}{\partial \dot{a}} + \dot{\beta}(a, \xi) \frac{\partial}{\partial \dot{\xi}},
\end{equation}
where \( \alpha(a, \xi) \) and \( \beta(a, \xi) \) are the components of the symmetry generator, and the condition for Noether symmetry is that the Lie derivative of the Lagrangian vanishes, i.e., \( \mathcal{L}_X \mathcal{L} = 0 \). Applying this condition to the point-like Lagrangian (\ref{eq:lagrangian}), we obtain a set of partial differential equations by collecting the coefficients of \( \dot{a}^2 \), \( \dot{\xi}^2 \), \( \dot{a} \dot{\xi} \), and the terms independent of velocities
\begin{equation}
\label{eq:noether_c1}
 g(\xi) \left( \alpha + 2 a \frac{\partial \alpha}{\partial a} \right) + a \left( \beta g'(\xi) + a \frac{\partial \beta}{\partial a} \right) = 0,
\end{equation}
\begin{equation}
 \label{eq:noether_c2}
\xi \left( 3 \alpha + 4 \xi \frac{\partial \alpha}{\partial \xi} \right) - a \left( \beta - 2 \xi \frac{\partial \beta}{\partial \xi} \right) = 0,
\end{equation}
\begin{equation}
\label{eq:noether_c3}
2\alpha + 2 g(\xi) \frac{\partial \alpha}{\partial \xi} + a \left( \frac{\partial \beta}{\partial \xi} +  \frac{\partial \alpha}{\partial a} + \frac{a}{2\xi} \frac{\partial \beta}{\partial a} \right)  = 0,
\end{equation}
\begin{equation}
\label{eq:noether_c4}
3 V(\xi) \alpha + a \beta V'(\xi) = 0.
\end{equation}
To solve these equations, we assume a potential of the form \( V(\xi) = V_0 \xi^n \), as specified in the model, and adopt the ansatz for the symmetry generator components
\begin{equation}
\label{eq:alpha_beta}
\alpha(a, \xi) = \alpha_0 a^k \xi^m, \quad \beta(a, \xi) = -\frac{3 \alpha_0}{n} a^{k-1} \xi^{m+1},
\end{equation}
where \( \alpha_0 \), \( k \), \( m \), and \( n \) are constants to be determined, and the form of \( \beta \) is chosen to satisfy the constraints from (\ref{eq:noether_c4}). We also assume \( g(\xi) = g_0 \xi \), a linear coupling function, to simplify the computations. Substituting these into (\ref{eq:noether_c1})--(\ref{eq:noether_c4}) and solving the resulting system, we obtain
\begin{align}
\label{eq:noether_solutions}
m = -\frac{3 (n - 1)}{2 (2n - 3)}, \quad k = \frac{-9 + 15n - 8n^2}{2n (2n - 3)}, \nonumber \\
 g_0 = \frac{3 (4n - 3)}{4 n^2}, \quad V(\xi) = V_0 \xi^n,
\end{align}
which implies:
\begin{equation}
\label{eq:f_xi}
g(\xi) = \frac{3 (4n - 3)}{4 n^2} \xi.
\end{equation}
The symmetry generator components are
\begin{equation}
\label{eq:alpha_solution}
\alpha(a, \xi) = \alpha_0 a^{\frac{-9 + 15n - 8n^2}{2n (2n - 3)}} \xi^{-\frac{3 (n - 1)}{2 (2n - 3)}},
\end{equation}
\begin{equation}
\label{eq:beta_solution}
\beta(a, \xi) = -\frac{3 \alpha_0}{n} a^{\frac{9 - 21n + 12n^2}{2n (3 - 2n)}} \xi^{\frac{n - 3}{4n - 6}}.
\end{equation}
These solutions satisfy the Noether symmetry conditions and constrain the functional forms of \( g(\xi) \) and \( V(\xi) \), enabling further simplification of the dynamical system. The value \( n = \frac{3}{2} \) is excluded, as this value leads to the trivial symmetry generator with non-physical results. The conserved quantity associated with the Noether symmetry will be used in the next subsection to introduce cyclic variables and derive exact solutions for \( a(t) \) and \( \xi(t) \).

\subsection{Cyclic Variable and Exact Solutions}
\label{subsec:cyclic_variables}

The Noether symmetry identified in the previous subsection allows us to introduce a cyclic variable that simplifies the Lagrangian and facilitates the derivation of exact solutions for the scale factor \( a(t) \) and the scalar field \( \xi(t) \). The coordinate transformation $(a,\xi)\to(s,u)$ generated via Noether symmetry, such that the new variable $s$ is cyclic, satisfies the following equations
\begin{equation}
\label{eq:cyclic_conditions}
\alpha(a, \xi) \frac{\partial s}{\partial a} + \beta(a, \xi) \frac{\partial s}{\partial \xi} = 1, \quad \alpha(a, \xi) \frac{\partial u}{\partial a} + \beta(a, \xi) \frac{\partial u}{\partial \xi} = 0.
\end{equation}
Using the symmetry generator components (\ref{eq:alpha_solution}) and (\ref{eq:beta_solution}), we solve these equations to obtain
\begin{equation}
\label{eq:cyclic_z}
s(a, \xi) = \frac{n a^{\frac{3 (3 - 7n + 4n^2)}{2n (2n - 3)}} \xi^{\frac{3 (n - 1)}{4n - 6}}}{3 (n - 1) \alpha_0},
\end{equation}
\begin{equation}
\label{eq:cyclic_u}
u(a, \xi) = \log \left( a^{\frac{3}{n}} \xi \right).
\end{equation}
where $n\neq 1$. Inverting these transformations, we express \( a \) and \( \xi \) in terms of \( u \) and \( s \)
\begin{equation}
\label{eq:a_solution}
a(t) = \left( \frac{3 (n - 1) \alpha_0 s(t)}{n} \right)^{\frac{n}{3 (n - 1)}} e^{\frac{n u(t)}{2 (3 - 2n)}},
\end{equation}
\begin{equation}
\label{eq:xi_solution}
\xi(t) = \left( \frac{3 (n - 1) \alpha_0 s(t)}{n} \right)^{\frac{-1}{n - 1}} e^{\frac{(4n - 3) u(t)}{4n - 6}}.
\end{equation}
Substituting these into the point-like Lagrangian (\ref{eq:lagrangian}), with \( g(\xi) = \frac{3 (4n - 3)}{4 n^2} \xi \) from Eq. (\ref{eq:f_xi}) and \( V(\xi) = V_0 \xi^n \), we obtain the transformed Lagrangian in terms of the cyclic variable
\begin{equation}
\label{eq:lagrangian_cyclic}
\mathcal{L} = \frac{\alpha_0 (2n - 3)}{4n} e^{\frac{n - 3}{2 (2n - 3)} u} \dot{u} \dot{s} - V_0 e^{n u}.
\end{equation}
The equations of motion are derived by varying Eq. (\ref{eq:lagrangian_cyclic}) with respect to \( u \) and \( s \). For \( s \), the Euler-Lagrange equation yields
\begin{equation}
\label{eq:eom_z}
\frac{\alpha_0 (2n - 3)}{4n} e^{\frac{n - 3}{2 (2n - 3)} u} \dot{u} = I_0,
\end{equation}
where \( I_0 \) is the conserved quantity associated with the Noether symmetry. For \( u \), the equation of motion is
\begin{equation}
\label{eq:eom_u}
-n V_0 e^{n u} - \frac{\alpha_0 (2n - 3)}{4n} e^{\frac{n - 3}{2 (2n - 3)} u} \ddot{s} = 0.
\end{equation}
The energy function, defined as \( E = \dot{u} \frac{\partial \mathcal{L}}{\partial \dot{u}} + \dot{s} \frac{\partial \mathcal{L}}{\partial \dot{s}} - \mathcal{L} \), is:
\begin{equation}
\label{eq:energy_cyclic}
E = V_0 e^{n u} + \frac{\alpha_0 (2n - 3)}{4n} e^{\frac{n - 3}{2 (2n - 3)} u} \dot{u} \dot{s}.
\end{equation}
Setting \( E = 0 \), consistent with the FRW spacetime, we obtain
\begin{equation}
\label{eq:energy_zero}
V_0 e^{n u} + \frac{\alpha_0 (2n - 3)}{4n} e^{\frac{n - 3}{2 (2n - 3)} u} \dot{u} \dot{s} = 0.
\end{equation}
Solving Eq. (\ref{eq:eom_z}) for \( \dot{u} \):
\begin{equation}
\label{eq:u_dot}
\dot{u} = \frac{4n I_0}{\alpha_0 (2n - 3)} e^{-\frac{n - 3}{2 (2n - 3)} u}.
\end{equation}
This differential equation has the solution as

\begin{equation}
\label{eq:u_t}
u(t) = \frac{2 (2n - 3)}{n - 3} \log \left( kt + u_0 \right),
\end{equation}
where \( u_0 \) is an integration constant, $n \neq 3$ and $k$ is defined by $k= \frac{2n (n-3) I_0 }{\alpha_0 (2n - 3)^2}$. Substituting Eq. (\ref{eq:u_t}) into Eq. (\ref{eq:energy_zero}) and solving for \( z(t) \), we obtain

\begin{equation}
\label{eq:z_t}
s(t) = -\frac{ V_0 (n-3)}{k I_0 (4n^2-5n-3)} \left(kt+u_0 \right)^{\frac{4n^2-5n-3}{n - 3}} + s_0,
\end{equation}
where \( s_0 \) is another integration constant. Finally, substituting Eqs. (\ref{eq:u_t}) and (\ref{eq:z_t}) into Eqs. (\ref{eq:a_solution}) and (\ref{eq:xi_solution}), we derive the exact solutions for the scale factor and the scalar field as

\begin{align}
\label{eq:a_t_final}
a(t) = \left( \frac{3 (n - 1) \alpha_0}{n} \right)^{\frac{n}{3 (n - 1)}}\left(kt+ u_0 \right)^{\frac{n}{3 - n}}  \nonumber \\
\left[-\frac{ V_0 (n-3)}{k I_0 (4n^2-5n-3)} \left(kt+u_0 \right)^{\frac{4n^2-5n-3}{n - 3}} + s_0, \right]^{\frac{n}{3 (n - 1)}} ,
\end{align}
\begin{align}
\label{eq:xi_t_final}
\xi(t) = \left( \frac{3 (n - 1) \alpha_0}{n} \right)^{\frac{1}{1-n}}\left( kt+ u_0 \right)^{\frac{4n - 3}{n - 3}} \nonumber \\
 \left[ -\frac{ V_0 (n-3)}{k I_0 (4n^2-5n-3)} \left(kt+u_0 \right)^{\frac{4n^2-5n-3}{n - 3}} + s_0 \right]^{\frac{1}{1-n }} .
\end{align}
We now examine the asymptotic behavior of the scale factor $a(t)$ given in Eq. (\ref{eq:a_t_final}) in the two limits: $t \to 0$ and $t \to \infty$. Assuming $u_0 > 0$, as $t \to 0$ we have
\begin{equation}
(kt + u_0) \to u_0, \quad so \quad (kt + u_0)^{powers} \to const.
\end{equation}
Thus, the dominant behavior comes from the additive constant inside the bracket
\begin{equation}
a(t) \approx const \times \left[ s_0 - \frac{V_0(n - 3)}{kI_0(4n^2 - 5n - 3)}u_0^{\frac{4n^2 - 5n - 3}{n - 3}} \right]^{\frac{n}{3(n - 1)}} u_0^{\frac{n}{3 - n}}.
\end{equation}
Hence, as \( t \to 0 \), \( a(t) \to const \). This indicates that the universe starts with a non-zero scale factor, which may correspond to a bouncing or emergent cosmological scenario, depending on the sign and magnitude of the constants. In the limit \( t \to \infty \), the power-law terms dominate. Ignoring the additive constant \( s_0 \) in comparison to the growing term, we approximate:
\begin{equation}
a(t) \sim \left( const \times (kt + u_0)^{\frac{4n^2 - 5n - 3}{n - 3}} \right)^{\frac{n}{3(n - 1)}}
(kt + u_0)^{\frac{n}{3 - n}}.
\end{equation}
Combining exponents, we obtain the leading behavior
\begin{equation}
a(t) \sim t^{\gamma_n}, \quad with \quad \gamma_n = \frac{n(4n^2 - 5n - 3)}{3(n - 1)(n - 3)} + \frac{n}{3 - n}.
\end{equation}
This shows that the scale factor follows a power-law expansion at late times, and the exponent $\gamma_n$ depends sensitively on the choice of $n$. If $\gamma_n > 1$, the expansion accelerates; otherwise, it decelerates. The general solution for $a(t)$ in the $n \neq 1,3$ case interpolates between a constant (or slowly varying) early-time behavior and a power-law expansion at late times. The acceleration or deceleration of the universe in this regime is controlled by the parameter $n$ through the effective exponent $\gamma_n$.
These solutions describe the evolution of the universe for a general \( n \neq 1, 3,\), with parameters \( \alpha_0 \), \( V_0 \), \( I_0 \), \( u_0 \), and \( s_0 \) determining the specific cosmological dynamics. The solutions can be further analyzed to determine the behavior of the scale factor and scalar field during different cosmological epochs, such as inflation or late-time acceleration, by choosing appropriate values of \( n \).

\section{Special Cases for \( n \)}
\label{sec:special_cases}
In this section, we address specific values of the parameter \( n \) that require separate analysis due to singularities or unique dynamical behaviors in the Noether symmetry approach. Specifically, we first focus on cases such as \( n=1 \), where the definitions of the coordinate transformation obtained from symmetry conditions, as derived in Section (\ref{sec:noether}), become ill-defined or lead to distinct solutions. The goal is to rederive the Lagrangian with cyclic variable. and obtain exact solutions for the scale factor \( a(t) \) and scalar field \( \xi(t) \), while also analyzing the cosmological implications through parameters such as the deceleration parameter and the equation of state parameter.

\subsection{Case \( n=1 \)}
\label{subsec:n_equals_1}
For the case \( n=1 \), the coordinate transformation \( s \) in (\ref{eq:cyclic_z}) becomes problematic because the denominator \( (n - 1) \) vanishes, rendering the expression undefined. This necessitates a re-evaluation of the Noether symmetry conditions and the coordinate transformation specifically for \( n=1 \). Substituting \( n=1 \) into the Noether symmetry solutions (\ref{eq:noether_solutions}), we obtain the following symmetry generator components and coupling function as
\begin{equation}
\label{eq:alpha_beta_n1}
\alpha(a, \xi) = \alpha_0 a, \quad \beta(a, \xi) = -3 \alpha_0 \xi,
\end{equation}
\begin{equation}
\label{eq:g_n1}
g(\xi) = \frac{3}{4} \xi, \quad V(\xi) = V_0 \xi,
\end{equation}
where \( g_0 = \frac{3}{4} \) and \( V(\xi) = V_0 \xi \) is the potential for \( n=1 \). These results are derived by re-solving the Noether symmetry conditions (\ref{eq:noether_c1})--(\ref{eq:noether_c4}) with \( n=1 \), ensuring consistency with the Lagrangian (\ref{eq:lagrangian}).

To proceed, we redefine the coordinate transformation \( u \) and \( s \) to satisfy the Noether symmetry conditions (\ref{eq:cyclic_conditions}). Solving these for \( n=1 \), we find as
\begin{equation}
\label{eq:cyclic_z_n1}
s(a, \xi) = \frac{\log a}{\alpha_0},
\end{equation}
\begin{equation}
\label{eq:cyclic_u_n1}
u(a, \xi) = a^3 \xi.
\end{equation}
Inverting these transformations, we express \( a(t) \) and \( \xi(t) \) as
\begin{equation}
\label{eq:a_xi_n1}
a(t) = e^{\alpha_0 s(t)}, \quad \xi(t) = u(t) e^{-3 \alpha_0 s(t)}.
\end{equation}
Substituting (\ref{eq:a_xi_n1}) into the point-like Lagrangian (\ref{eq:lagrangian}), with \( g(\xi) = \frac{3}{4} \xi \) and \( V(\xi) = V_0 \xi \), we obtain the transformed Lagrangian in terms of \( u \) and \( s \) as follows
\begin{equation}
\label{eq:lagrangian_n1}
\mathcal{L} = \frac{1}{8} \left( -8 V_0 u + \frac{\dot{u}^2}{u} - 2 \alpha_0 \dot{u} \dot{s} \right).
\end{equation}
The equations of motion are derived by varying Eq. (\ref{eq:lagrangian_n1}) with respect to \( s \) and \( u \). For the variable \( s \), the Euler-Lagrange equation yields
\begin{equation}
\label{eq:eom_z_n1}
-\frac{1}{4} \alpha_0 \dot{u} = I_0,
\end{equation}
where \( I_0 \) is the conserved quantity associated with the Noether symmetry. For the variable \( u \), the equation of motion is
\begin{equation}
\label{eq:eom_u_n1}
-8 V_0 + \frac{\dot{u}^2 - 2 u \ddot{u}}{u^2} + 2 \alpha_0 \ddot{s} = 0.
\end{equation}
The energy function, defined as \( E = \dot{u} \frac{\partial \mathcal{L}}{\partial \dot{u}} + \dot{s} \frac{\partial \mathcal{L}}{\partial \dot{s}} - \mathcal{L} \), is
\begin{equation}
\label{eq:energy_n1}
E = V_0 u + \frac{\dot{u}^2}{8 u} - \frac{1}{4} \alpha_0 \dot{u} \dot{s}.
\end{equation}
Setting \( E = 0 \), we have
\begin{equation}
\label{eq:energy_zero_n1}
V_0 u + \frac{\dot{u}^2}{8 u} - \frac{1}{4} \alpha_0 \dot{u} \dot{s} = 0.
\end{equation}
Solving (\ref{eq:eom_z_n1}) for \( \dot{u} \):
\begin{equation}
\label{eq:u_dot_n1}
\dot{u} = -\frac{4 I_0}{\alpha_0}.
\end{equation}
Integrating, we find
\begin{equation}
\label{eq:u_t_n1}
u(t) = -\frac{4 I_0}{\alpha_0} t + u_0,
\end{equation}
where \( u_0 \) is an integration constant. Substituting (\ref{eq:u_t_n1}) into (\ref{eq:energy_zero_n1}) and solving for \( s(t) \), we obtain
\begin{equation}
\label{eq:z_t_n1}
s(t) = \frac{2 V_0}{\alpha_0} t^2-\frac{V_0 u_0}{I_0} t  + \frac{1}{2 \alpha_0} \log \left( -\frac{4 I_0 t}{\alpha_0} + u_0 \right) + s_0,
\end{equation}
where \( s_0 \) is another integration constant. Substituting (\ref{eq:u_t_n1}) and (\ref{eq:z_t_n1}) into (\ref{eq:a_xi_n1}), we derive the exact solutions
\begin{equation}
\label{eq:a_t_n1}
a(t) = \left(-\frac{4 I_0 t}{\alpha_0} + u_0\right)^{\frac{1}{2}}e^{\alpha_0 \left(\frac{2 V_0}{\alpha_0} t^2-\frac{V_0 u_0}{I_0} t  + s_0\right)} ,
\end{equation}
\begin{equation}
\label{eq:xi_t_n1}
\xi(t) =  \left(-\frac{4 I_0 t}{\alpha_0} + u_0\right)^{-\frac{1}{2}}e^{-3\alpha_0 \left(\frac{2 V_0}{\alpha_0} t^2-\frac{V_0 u_0}{I_0} t  + s_0\right)}.
\end{equation}
The Hubble parameter is
\begin{equation}
\label{eq:hubble_n1}
H(t) = \frac{\dot{a}}{a} = \frac{2I_0}{4I_0 t-u_0\alpha0}+ 4 V_0 t-\frac{u_0\alpha_0V_0}{I_0}.
\end{equation}
The analysis of these solutions in their current form can be quite complicated due to the numerous arbitrary constants they contain. So, we first assume that $I_{0}=-\alpha_{0}$. Then, following the steps in Ref. \cite{Demianski:2004qt,Rubano:2003et}, we set the initial condition that the scale factor is zero at $t = 0$, i.e., $a(0) = 0$, to ensure physical consistency; this implies $u_{0} = 0$ from Eq. (\ref{eq:a_t_n1}). As a second step, we take the present time as $t_{0}=1$. Thus, as a standard, we can assume the condition $a(1)=1$. This condition gives us $s_0=-\frac{2V_0+\log{2}}{\alpha_0}$. As a final condition, we assume that the present value of the Hubble parameter is $h_{0}$, which is not the same as the value of the Hubble constant obtained from observations. Applying this condition to Eq. (\ref{eq:hubble_n1}), we obtain $V_0=\frac{h_0}{4}-\frac{1}{8}$. After all these restrictions on the constants, the scale factor, the scalar field, the Hubble parameter, the deceleration parameter, and the equation of state parameter can be written as follows, depending only on the constant $h_0$
\begin{equation}
\label{scalef}
a(t) = t^{\frac{1}{2}}e^{\frac{(2h_0-1)}{4}(t^2-1)},
\end{equation}
\begin{equation}
\label{scalar}
\xi(t) = 4t^{-\frac{1}{2}}e^{-\frac{3(2h_0-1)}{4}(t^2-1)},
\end{equation}
\begin{equation}
\label{hub}
H(t) =\frac{1}{2t}+\left(h_0-\frac{1}{2}\right)t,
\end{equation}
\begin{equation}
\label{decel}
q(t) =-1+\frac{2\left[1-(2h_0-1)t^2\right]}{\left[1+(2h_0-1)t^2\right]^2},
\end{equation}
\begin{equation}
\label{eos}
\omega_\xi(t) =-1+\frac{4\left[1-(2h_0-1)t^2\right]}{3\left[1+(2h_0-1)t^2\right]^2},
\end{equation}
\begin{figure}
\resizebox{0.48\textwidth}{!}{\includegraphics{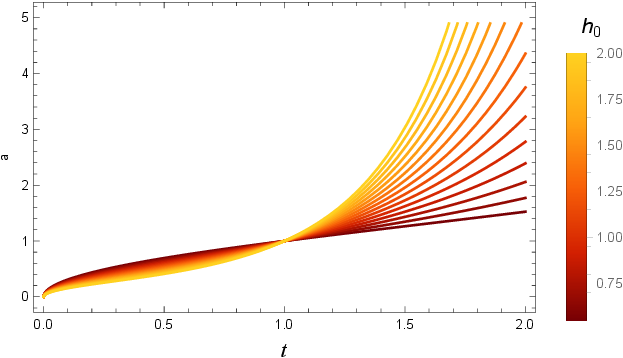}}
\caption{Evolution of the scale factor versus the time $t$ for the value $n=1$.} \label{g1}
\end{figure}
\begin{figure}
\resizebox{0.48\textwidth}{!}{\includegraphics{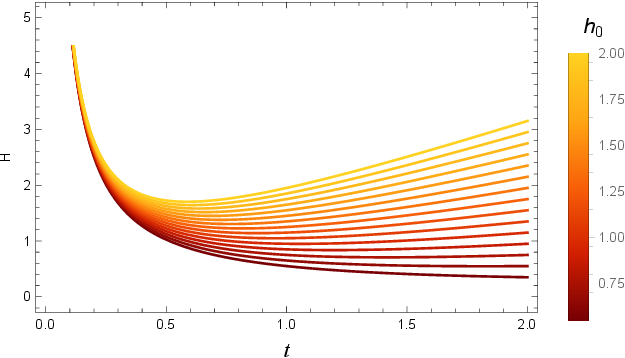}}
\caption{Evolution of the Hubble parameter versus the time $t$ for the value $n=1$.} \label{g2}
\end{figure}
\begin{figure}
\resizebox{0.48\textwidth}{!}{\includegraphics{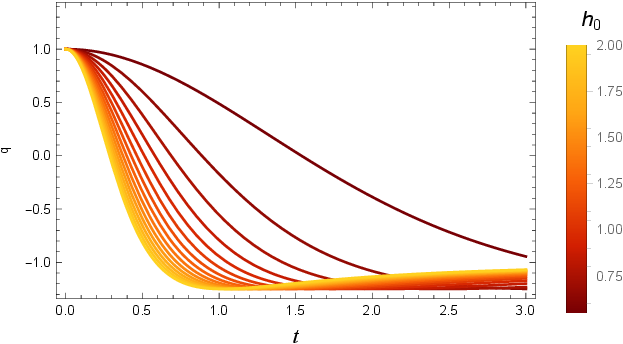}}
\caption{Evolution of the deceleration parameter versus the time $t$ for the value $n=1$.} \label{g3}
\end{figure}
\begin{figure}
\resizebox{0.48\textwidth}{!}{\includegraphics{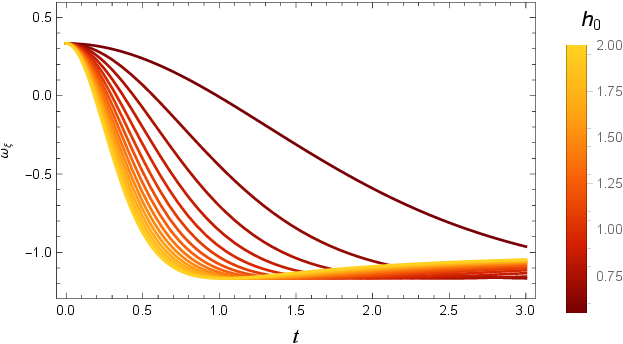}}
\caption{Evolution of the equation of state parameter versus the time $t$ for the value $n=1$.}\label{g4}
\end{figure}
\begin{figure}
\resizebox{0.48\textwidth}{!}{\includegraphics{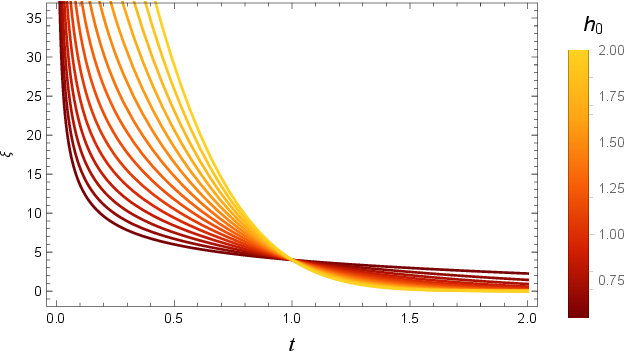}}
\caption{Evolution of the scalar field $\xi$ versus the time $t$ for the value $n=1$.} \label{g5}
\end{figure}
From these equations, we can easily conclude that for $h_{0}=\frac{1}{2}$, the model leads to a radiation-dominated universe. The evolution of these obtained cosmological parameters with time for different values of $h_{0}$ in the range $(0.5, 2]$ is shown in Figs. (\ref{g1})$-$ (\ref{g5}). In the early universe, the scale factor is extremely small while the Hubble parameter tends to infinity, implying that the expansion rate of the universe at the initial time is infinite. The scale factor increases monotonically with time, but the Hubble parameter decreases for a short time and then tends to increase in the future, such that the expansion rate increases more as the value of $h_{0}$ increases (See Figs. (\ref{g1}) and (\ref{g2})). For small \( t \), \( q(t) \approx 1 \), indicating a decelerating universe. However, from Fig. (\ref{g3}) we can say that the deceleration parameter becomes negative, suggesting an accelerating expanding universe. Also, for very large $t$, the deceleration parameter goes to $-1$. From Fig. (\ref{g4}), it is clear that the equation of state parameter $\omega_\xi$ exhibits a transition behavior from a dark energy-dominated universe with quintessence phase to a universe with phantom phase for large values of $h_{0}$. However, as the values of $h_{0}$ become smaller (but larger than $1/2$), the model shows a universe with only quintessence phases. Although the model parameter $h_{0}$ can be constrained by observational data, for example, for a value of $h_{0}=0.74$, we calculate the value of the deceleration parameter at present time as $q=-0.525$ and the value of the equation of state parameter at present time as $\omega_{\xi}=-0.683$. These calculated values are within the range of cosmological observations \cite{DESI:2025zgx,Pourojaghi:2024bxa}. The time evolution of the scalar quantity $\xi$ is given in Fig. (\ref{g5}). From this figure, it is seen that it becomes infinite in the limit $t\rightarrow 0$ and then decreases with time, finally going to zero in the limit $t\rightarrow\infty$. These results suggest that the \( n=1 \) case can describe a universe changing from deceleration to acceleration, with the scalar field \( \xi(t) \) playing a role analogous to the dark energy component.

\subsection{Case $n=3$}
In the general analysis for arbitrary $n$, we found that $n=3$ causes the new variable $u$ to be undefined. Therefore, in this subsection, we examine this case separately. We begin with the point-like Lagrangian for $n=3$, obtained by substituting $g(\xi) = \frac{3}{4} \xi$ and $V(\xi) = V_0 \xi^3$ into Eq. (\ref{eq:lagrangian}) and transforming to the new variables $u$ and $z$
\begin{equation} \label{lag1}
\mathcal{L} = \frac{1}{4} \alpha_0 \dot{u} \dot{z} - V_0 e^{3u}.
\end{equation}
This Lagrangian leverages the cyclic nature of $s$, simplifying the system for deriving analytical solutions.

\begin{figure}
\resizebox{0.48\textwidth}{!}{\includegraphics{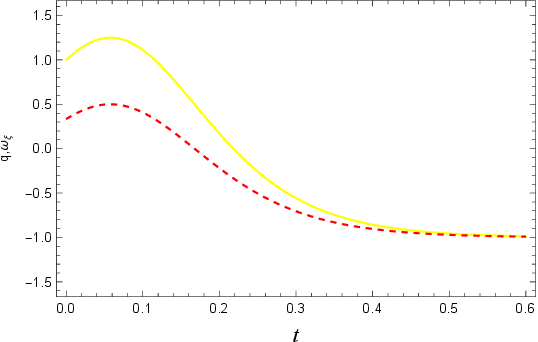}}
\caption{Plots of the deceleration parameter (yellow line) and the equation of state parameter (red line) versus the time $t$ for the value of $n = 3$.} \label{g6}
\end{figure}

Varying the Lagrangian (\ref{lag1}) with respect to $s$, the Euler-Lagrange equation yields
\begin{equation} \label{lag2}
\frac{1}{4} \alpha_0 \dot{u} = I_0,
\end{equation}
where $I_0$ is the conserved quantity associated with the Noether symmetry. By integrating, we find
\begin{equation} \label{lag3}
u(t) = \frac{4 I_0}{\alpha_0} t + u_0,
\end{equation}
where $u_0$ is an integration constant. The energy function is:
\begin{equation} \label{lag4}
E = V_0 e^{3u} + \frac{1}{4} \alpha_0 \dot{u} \dot{s}.
\end{equation}
Setting $E = 0$, we have:
\begin{equation} \label{lag5}
V_0 e^{3u} + \frac{1}{4} \alpha_0 \dot{u} \dot{s} = 0.
\end{equation}
Substituting $u(t)$, we solve for $s$:
\begin{equation} \label{lag6}
s(t) = -\frac{V_0 \alpha_0}{12 I_0^2} e^{\frac{12 I_0 t}{\alpha_0} + 3 u_0} +s_0,
\end{equation}
where $s_0$ is a redefined integration constant. The coordinate transformation relate to $a(t)$ and $\xi(t)$ via:
\begin{equation}\label{lag7}
a(t) = e^{-\frac{u(t)}{2}}\left(2\alpha_0 s(t)\right)^{\frac{1}{2}} , \\
\quad \xi(t) =e^{\frac{3u(t)}{2}}\left(2\alpha_0 s(t) \right)^{-\frac{1}{2}}.
\end{equation}
Substituting $u(t)$ and $s(t)$ from (\ref{lag3}) and (\ref{lag6}), we obtain
\begin{equation}\label{lag8}
a(t) = \left[2 s_0 \alpha_0 e^{-\left(\frac{4I_0 t}{\alpha_0}+u_0\right)}-\frac{V_0\alpha_0^2}{6I_0^2}e^{2\left(\frac{4I_0 t}{\alpha_0}+u_0\right)}\right]^{\frac{1}{2}},
\end{equation}
\begin{equation}\label{lag9}
\xi(t) = \left[2 s_0 \alpha_0 e^{-3\left(\frac{4I_0 t}{\alpha_0}+u_0\right)}-\frac{V_0\alpha_0^2}{6I_0^2}\right]^{-\frac{1}{2}}.
\end{equation}
To ensure physical consistency, we impose the boundary condition $a(0) = 0$. From (57), at $t=0$, Solving for $s_0$ we find
\begin{equation}\label{lag10}
s_0 = \frac{V_0 \alpha_0}{12 I_0^2} e^{3 u_0}.
\end{equation}
Substituting $s_0$ into Eq. (\ref{lag8}), the scale factor becomes
\begin{equation}\label{lag11}
a(t) = \left[\frac{V_0 \alpha_0^2e^{2u_0}}{6I_0^2}\left(e^{-\frac{4I_0 t}{\alpha_0}}-e^{\frac{8I_0 t}{\alpha_0}}\right)\right]^{\frac{1}{2}}.
\end{equation}
The deceleration parameter, defined as $q(t) = -\frac{\ddot{a} a}{\dot{a}^2}$, is computed using Eq. (\ref{lag11})
\begin{equation}\label{lag12}
q(t) =-1+\frac{18e^{\frac{12 I_0 t}{\alpha_0}}}{\left( 1 + 2 e^{\frac{12 I_0 t}{\alpha_0}} \right)^2} .
\end{equation}\label{lag13}
The equation of state parameter is calculated for this case as
\begin{equation}
\omega_{\xi}(t) =-1+\frac{12e^{\frac{12 I_0 t}{\alpha_0}}}{\left( 1 + 2 e^{\frac{12 I_0 t}{\alpha_0}} \right)^2} .
\end{equation}
The results of the evolution of the deceleration parameter and the equation of state parameter with time are shown in Fig. (\ref{g6}), where $I_{0}=-\alpha_{0}$ is taken for convenience. When we look at the evolution of the deceleration parameter in Fig. (\ref{g6}), we see that there is a transition from a decelerating expanding universe to an accelerating expanding universe, similar to the case of $n=1$. This model depicts a universe that is perpetually accelerating in the late epoch, with the vector field $\xi(t)$ acting as a dominant dark energy-like component throughout its evolution. However, unlike the case of $n=1$, which mimics the transition from the quintessence phase to the phantom phase, such a transition never occurs in the case of $n=3$, and the phantom dividing line is not crossed. As a result, the $n = 3$ model may fully match the observed cosmological history, during periods when early matter or radiation dominated, and especially during the periods when dark energy dominated. Further validation against observational data, such as supernova light curves or cosmic microwave background measurements, is required to assess its physical relevance.
\section{Conclusion}
\label{sec:conclusion}
In this study, a cosmological model was investigated in the framework of teleparallel gravity, in which the vector field $A_\mu$ is non-minimally coupled to the torsion scalar $T$ via a function of $\xi = A_\mu A^\mu$. In order to analyze the dynamic behavior of this model, the Noether symmetry approach was used. This approach determined the permissible forms for the coupling function $g(\xi)$ and the potential $V(\xi) = V_0 \xi^n$ and provided the possibility of finding exact and analytical solutions for the scale factor $a(t)$ and the field $\xi(t)$ through the coordinate transformation obtained, in particular, by using the symmetry vector. For the general form of the potential $V(\xi) = V_{0}\xi^n$, the Noether symmetry condition allowed the coupling function to take a linear form depending on the parameter $n$. For this general case, the coordinate transformation simplified the field equations of the considered model, yielding analytical solutions for the scale factor $a(t)$ and the field $\xi(t)$, as in Eqs. (\ref{eq:a_t_final}) and (\ref{eq:xi_t_final}, respectively. When the late-time behavior of the scale factor is examined, it is observed that it is in power-law form and that, depending on the values of the parameter $n$, the model can lead to an accelerating expanding universe.\\
In examining the general case for arbitrary values of the parameter $n$, it was observed that for some special values such as $n=1$ and $n=3$, a separate analysis is required, because in these cases, the dynamic behavior of the model or the mathematical structure of the equations becomes different. In the case $n=1$, the Noether symmetry approach restricts both the coupling function and the potential to be in linear form. Some analytically obtained cosmological parameters for this model, which depend only on the value of $h_{0}$, are given explicitly in Eqs. (\ref{scalef})-(\ref{eos}). From these parameters, it is clear that the model points to a radiation-dominated universe in the early stages. In addition, the values at present-day time, that is, at $t_{0}=1$, also depend on the $h_{0}$ parameter and their evolution over time is given in Figs. (\ref{g1})-(\ref{g5}). From these figures, we observe that the universe transitions from a decelerating phase to a phase with positive acceleration, and the equation of state parameter tends to a value close to $-1$ at late times. Also, the model can enter the phantom phase at some parameter values.\\
In contrast, in the $n=3$ case, the cosmological model first goes through a deceleration phase and then enters an acceleration phase, with the difference that the transition to the phantom phase is not observed in this case. These features indicate that the model can describe a diverse range of cosmological scenarios depending on the value of the potential energy. Overall, this research shows that the use of vector fields in teleparallel gravity, along with the use of Noether symmetry, can provide an effective framework for building analytical cosmological models that are comparable to observations.


\end{document}